\documentclass[12pt]{article}
\usepackage{amssymb,amsmath,graphicx,epsfig} 
\def\r{\rightarrow}
\def\ksb {\overline{{K^0}^*}}
\def\kst {{K^0}^*}
\def\issue(#1,#2,#3){{\bf #1}, #2 (#3)} % AIP format

\def\APP(#1,#2,#3){{\rm Acta Phys.\ Polon.} \ \issue({\bf #1},#2,#3)}
\def\ANP(#1,#2,#3){{\rm Annals of Physics} \ \issue({\bf #1},#2,#3)}
\def\ARNPS(#1,#2,#3){{\rm Ann.\ Rev.\ Nucl.\ Part.\ Sci.} \ \issue({\bf #1},#2,#3)}
\def\CPC(#1,#2,#3){{\rm Comp.\ Phys.\ Comm.} \ \issue({\bf #1},#2,#3)}
\def\CIP(#1,#2,#3){{\rm Comput.\ Phys.} \ \issue({\bf #1},#2,#3)}
\def\EPJ(#1,#2,#3){{\rm Eur.\ Phys.\ J.} \ \issue({\bf #1},#2,#3)}
\def\EPJD(#1,#2,#3){Eur.\ Phys.\ J. Direct\ C \ \issue({\bf #1},#2,#3)}
\def\IJMP(#1,#2,#3){{\rm Int.\ J.\ Mod.\ Phys.} \ \issue({\bf #1},#2,#3)}
\def\JHEP(#1,#2,#3){{\rm J.\ High Energy Physics} \ \issue({\bf #1},#2,#3)}
\def\JP(#1,#2,#3){{ J.\ Phys.} \ \issue({\bf #1},#2,#3)}
\def\MPL(#1,#2,#3){{Mod.\ Phys.\ Lett.} \ \issue({\bf #1},#2,#3)}
\def\NP(#1,#2,#3){{Nucl.\ Phys.} \ \issue({\bf #1},#2,#3)}
\def\NIM(#1,#2,#3){{ Nucl.\ Instrum.\ Meth.} \ \issue({\bf #1},#2,#3)}
\def\PL(#1,#2,#3){{ Phys.\ Lett.} \ \issue({\bf #1},#2,#3)}
\def\PR(#1,#2,#3){{ Phys.\ Rev.} \ \issue({\bf #1},#2,#3)}
\def\PRL(#1,#2,#3){{ Phys.\ Rev.\ Lett.} \ \issue({\bf #1},#2,#3)}
\def\SJNP(#1,#2,#3){{ Sov.\ J. Nucl.\ Phys.} \ \issue({\bf #1},#2,#3)}
\def\ZP(#1,#2,#3){{Zeit.\ Phys.} \ \issue({\bf #1},#2,#3)}
\def\be {\begin{equation}}
\def\ee {\end{equation}}
\def\bea {\begin{eqnarray}}
\def\eea {\end{eqnarray}}
\def\rpv {{R_p}\!\!\!/}
\begin{document}
\begin{titlepage}
%\tableofcontents
%\pagestyle{empty}
%\begin{flushright}
%CU-PHYSICS/x-2010
%\end{flushright}
\begin{center}
{\Large {\bf 
{Rare $\tau$ Decays in R-parity Violating Supersymmetry}}}\\[5mm]
\bigskip
{\sf Roshni Bose} 
%and {\sf Anirban Kundu}

\bigskip

{\footnotesize\rm 
Department of Physics, University of Calcutta, \\
92, Acharya Prafulla Chandra Road, Kolkata 700 009, India}

\normalsize
\vskip 10pt

{\large\bf Abstract}
\end{center}

\begin{quotation} \noindent 

We constrain, from rare $\tau$ decays,
 several combinations of $\lambda$ and $\lambda'$ type
couplings coming from Supersymmetry without R-parity. 
The processes that we consider are $\tau \to \ell M$, 
$\tau\to \ell_i \ell_j \ell_k$, and $\tau\to \ell\gamma$,
where $\ell$ stands for either $e$ or $\mu$, and $M$
is the generic symbol for a meson. We update several existing
bounds, and provide a few new ones too.

\vskip 10pt
PACS numbers: {\tt 12.60.Jv, 13.35.Dx}\\
\end{quotation}
\begin{flushleft}\today\end{flushleft}
\vfill
\end{titlepage}
%%%%%%%%%%%%%%%%%%%%%%%%%%%%%%%%%%%%%%%%%%%%%%%%%%%%%%%%%%%%%%%%%%%%%%%%%%%%%%
\newpage
\setcounter{page}{1}
%\centerline{\large\sc 1. Introduction}
\section{Introduction}
\label{sec1}

The hope to discover physics beyond the Standard Model (SM), which we might
call New Physics (NP) for simplicity, is the principal {\em raison d'\^{e}tre} for
particle physicists. There are several motivations, and there are innumerable
candidates, but Supersymmetry, in its various avatars, stand out. As it appears,
lepton and baryon numbers, L and B respectively, are good symmetries of
the minimal Supersymmetric SM, but they are accidental symmetries in the
sense that L and B violating interactions are not {\em a priori} forbidden
in the superpotential, and one has to impose this {\em ad hoc} for B and L
violating couplings to be zero at every energy scale. What one does is to
consider a discrete symmetry, called R-parity, defined as 
\be
R_p = (-1)^{3{\rm B+L+2S} }\,,
\ee
where S is the spin of the particle. By definition, $R_p=+1$ for particles and
$R_p=-1$ for superparticles, and $R_p$ is imposed as a good symmetry of
the superpotential. This forbids B and L violating interactions separately, 
and makes the lightest supersymmetric particle (LSP) stable and a good
candidate for the dark matter. On the other hand, if $R_p$ is not a good symmetry,
the signatures change drastically, because all superparticles, including the LSP,
can decay inside the detector. 

There can be 45 $R_p$-violating (RPV) couplings in the superpotential
coming from the renormalizable terms
\be
W_{\rpv} = \lambda_{ijk} L_i L_j E^c_k + \lambda'_{ijk} L_i Q_j D^c_k + 
\lambda''_{ijk} U^c_i D^c_j D^c_k\,,
\label{wrpv}
\ee
where $L$, $Q$, $E$, $U$ and $D$ stand for lepton doublet, quark doublet,
lepton singlet, up-type quark singlet, and down-type quark singlet superfields
respectively; $i,j,k$ are generation indices that can run from 1 to 3; and 
$\lambda_{ijk}$ ($\lambda''_{ijk}$) are antisymmetric in $i$ and $j$ ($j$
and $k$). The phenomenology of RPV supersymmetry, including the collider
signatures and bounds on these couplings, may be found in \cite{barbier}.

In this work we would like to constrain several L-violating coupling 
combinations from rare $\tau$ decays.  
The decays that we will consider are $\tau\to \ell M$, $\tau\to \ell_i
\ell_j \ell_k$, and $\tau\to \ell \gamma$, where $\ell$ stands for 
$e$ or $\mu$, $M$ for any generic meson, and $i,j,k$ are generation
indices that can be 1 or 2. These decays being L violating, we will 
consider only $\lambda$ and $\lambda'$ type couplings, as simultaneous
L and B violation would lead to an unacceptably fast proton decay.  
There are some such studies in the literature 
\cite{tau1,tau3,tau4} whose bounds we will update.
We will also give a number of new bounds coming from radiative $\tau$
decays. However, as we will see, most of the combinations are also
bound from leptonic and semileptonic lepton flavour violating (LFV)
decays, and those numbers are better by more than an order of
magnitude.

The paper is organized as follows. 
In Section II, we compile the relevant expressions. Section III deals with
the analysis and the bounds that we obtain. Section IV is on the possible
implications of these bounds on the ongoing experiments. We summarize and
conclude in Section V.

%%%%%%%%%%%%%%%%%%%%%%%%%%%%%%%%%%%%%%%%%%%%%%%%%%%%
\section{Relevant Expressions}
%%%%%%%%%%%%%%%%%%%%%%%%%%%%%%%%%%%%%%%%%%%%%%%%%%%%

We work in the framework of an explicit RPV model, with a hierarchical 
scheme of couplings: only those couplings that are relevant will be assumed
to dominate the others. For all decays that we will be interested in, we
need two different couplings, of the form $\lambda\lambda$, $\lambda\lambda'$,
or $\lambda'\lambda'$. We assume the hierarchical structure at the weak scale,
and do not investigate the possible high-scale theory behind them. Also, at the
weak scale, we assume these terms to appear in the mass basis of the quarks,
and not in the flavour basis, so that there is no further constraints on other couplings
coming from a Cabibbo-Kobayashi-Maskawa (CKM) type rotation.
We also
assume the bilinear couplings of the generic form $-\mu_i L_i H_2$, where
$H_2$ is the superfield that gives mass to charged leptons and down-type
quarks, to be zero at the weak scale. This also relaxes the possible constraints
coming from the neutrino masses and mixing angles in presence of the bilinear
terms. However, even some trilinear combinations like $\lambda^{(')}_{ikl}
\lambda^{(')}_{jlk}$ can generate nonzero entries for the $ij$-th element of
the neutrino mass ${\cal M}_\nu$ \cite{barbier}.

For the decays $\tau\to \ell M$ or $\tau\to 3\ell$, the relevant four-fermion
operator, which one may get by integrating the sfermion field out in eq.\ 
(\ref{wrpv}). The expression reads \cite{tau3,tau4}
\bea
{\cal H}_{\rpv} &=& 
L_{jklm} (\bar\ell_l(1+\gamma_5) \ell_m) (\bar\ell_k (1-\gamma_5) \ell_j)
\nonumber\\
&& + A_{jklm} (\bar\ell_j (1-\gamma_5) \ell_k) (\bar d_m (1+\gamma_5) d_l)
\nonumber\\
&& - \frac12 B_{jklm} (\bar\ell_j \gamma^\mu(1-\gamma_5) \ell_l) (\bar d_m \gamma_\mu(1+\gamma_5) d_k)
\nonumber\\
&& + \frac12 C_{jklm} (\bar\ell_j \gamma^\mu(1-\gamma_5) \ell_l) (\bar u_k \gamma_\mu(1-\gamma_5) u_m) + {\rm H.c.}\,,
\label{4fer}
\eea
where
\bea
&&
L_{jklm} = \sum_{i=1}^3 \frac{\lambda_{ijk}\lambda^\ast_{ilm}}{4M_{\tilde\nu_{iL}}^2}\,,\ \ 
A_{jklm} = \sum_{i=1}^3 \frac{\lambda^\ast_{ijk}\lambda'_{ilm}}{4M_{\tilde\nu_{iL}}^2}\,,\nonumber\\
&&
B_{jklm} = \sum_{i=1}^3 \frac{\lambda'_{lim}{\lambda'}^\ast_{jik}}{4M_{\tilde  u_{iL}}^2}\,,\ \ 
C_{jklm} = \sum_{i=1}^3 \frac{\lambda'_{lmi}{\lambda'}^\ast_{jki}}{4M_{\tilde  d_{iR}}^2}\,.
\label{labc}
\eea
We take, as a benchmark value, all sfermions to be degenerate at 100 GeV. 
While this value is experimentally not favoured any more, the bounds scale 
with $M_{\tilde f}^2$, as is evident from
eqs.\ (\ref{4fer}) and (\ref{labc}), and it is easier to compare with earlier results
available in the literature. 

If the final state consists of three leptons, only the $L$ term is important.
For a final state with a meson consisting of two down-type quarks, both $A$
and $B$ terms contribute. If the meson consists of up-type quarks only, 
the $C$ term is relevant. For mesons like $\pi^0$, $\eta$, or $\eta'$ in the
final state, all terms $A,B,C$ are important. Anyway, we will consider only
one of them to be nonzero at a time, so we do not entertain any possibility
of interference. There is, of course, no SM contributions to these processes.

We do not consider the running of the RPV couplings between the sfermion 
scale and the low-energy scale. The corrections, for a hierarchical scheme,
consists of a scaling by a factor between 2 and 3 \cite{dreiner}, and hence
can be dumped into the low-energy value of these couplings. The only QCD 
correction may occur between two quark fields; for $\tau$ decays, the pair
hadronizes, absorbing all the uncertainty in the decay constant. 

For mesonic decays of the $\tau$, the nonzero $\lambda\lambda'$ type 
couplings can only result in a pseudoscalar meson in the final state; that
a vector meson is forbidden is evident from the Lorentz structure of the $B$
and $C$ type operators. The decay width for a final state pseudoscalar
meson $P$ with quark content $\bar{q}_j q_k$, mass $M_P$, and decay
constant $f_P$, can be written as
\be
\Gamma(\tau\to P\ell_i) = \frac{(M_\tau^2+M_{\ell_i}^2-M_P^2) 
C(M_\tau^2, M_{\ell_i}^2, M_P^2) f_P^2 M_P^4}
{128\pi M_\tau^3 M_{\tilde{\nu}}^4 \left(M_{q_j}+M_{q_k}\right)^2)} 
|\lambda_{ni3}^\ast\lambda'_{njk}|^2\,,
\ee
where
\be
C(x,y,z) = \sqrt{x^2+y^2+z^2-2xy-2yz-2zx}\,.
\label{c-def}
\ee
The above expression has to be multiplied by $\frac12$ if there is a $\pi^0$
in the final state. Note that the combination $|\lambda_{n3i}^\ast \lambda'_
{nkj}|$ can also be bounded from the same decays, with identical expressions.

The generic couplings $B$ and $C$ mediate the decay $\tau\to\ell_i+M$
where $M$ is any pseudoscalar or vector meson, with quark contenti $\bar{q}_j
q_k$, and decay constant $f_M$. The decay widths are given by
\be
\Gamma(\tau\to\ell_i+M) =
\frac{f_M^2 C(M_\tau, M_{\ell_i}, M_M) F_0(M_\tau,M_{\ell_i},M_M)}
{512\pi {\tilde{M}}^4 M_\tau^3} |\lambda'_{3nk}{\lambda'}^\ast_{inj}|^2\,,
\ee
where the $C$-function is given in eq.\ (\ref{c-def}), $\tilde{M}$ is the
mass of the mediating sfermion, and
\bea
{\rm Pseudoscalar}~ &:&~F_0(x,y,z) = (x^2-y^2)^2 - z^2(x^2+y^2)\,,\nonumber\\
{\rm Vector}~ &:&~F_0(x,y,z) = z^2(x^2+y^2-z^2) + (x^2-y^2)^2 -z^4\,.
\eea
    
For the radiative decays, the invariant amplitude for $\tau\to\ell_i+\gamma$
can be written as \cite{tau1}
\bea
{\cal M} &=& i \bar{u}_\ell(k) T(1+\gamma_5) \sigma_{\mu\nu} q^\nu u_\tau(p)
\epsilon^\mu (q)\,,\nonumber\\
&=& T \bar{u}_\ell(k) (1+\gamma_5) \left( 2\epsilon.p - M_\tau \epsilon\!\!\!/~\right) 
u_\tau(p)\,,
\label{Mrad}
\eea
where $p$, $k$, and $q=p-k$ are the momenta of the $\tau$, the daughter lepton,
and the photon respectively. $\epsilon^\mu(q)$ is the polarization vector for
the photon, and $\sigma_{\mu\nu} = \frac{i}{2}[\gamma_\mu,\gamma_\nu]$. To get
the last line of the Gordon decomposition, we have neglected the final state 
lepton mass.
For $\lambda$-type couplings,
\be
T=\frac{ie\lambda_1\lambda_2M_{\tau}}{16\pi^2 {\tilde{M}}^2} F(x),
\ee
with
\be
F(x)=\frac{N_c}{24(x-1)^3}\left[\left(2x^2+5x-1-\frac{6x^2\ln x}{x-1}\right)-
\left(x^2-5x-2+\frac{6x \ln x}{x-1}\right)\right],
\ee
where $x$ is the ratio of fermion and sfermion masses squared: $x=M_f^2 / \tilde{M}^2$.
For $\lambda$-type couplings, they are leptons and sleptons, and $N_c=1$. For $\lambda'$-type
couplings, they are quarks and squarks, and $N_c=3$. One can neglect all fermion masses 
compared with the sfermions, except for the top quark, and in the limit $x\to 0$, 
$F(x) \to -N_c/24$. The expression is  simplified from the fact that  (s)neutrinos do not couple
to the photon. For $\lambda'$-type couplings, both up- and down-type (s)quarks may couple to
the photon, and the corresponding expression is more complicated \cite{tau1}:
\be
T = \frac{ie\lambda'_{1}\lambda'_{2}M_{\tau}}{16\pi^2 {\tilde M}^2}
\left[ F_1 \left(
\frac{M_d^2}{\tilde{M}^2}\right) + F_2 \left(\frac{M_u^2}{\tilde{M}^2}\right)\right]\,,
\ee
where
\bea
F_1(x)&=&\frac{N_c}{24(x-1)^3}\left[
\frac{2}{3}\left(2x^2+5x-1-\frac{6x^2\ln x}{x-1}\right)-\frac{1}{3}\left(x^2-5x-2+\frac{6x\ln x}{x-1}
\right)\right]\,,\nonumber\\
F_2(x)&=&\frac{N_c}{24(x-1)^3}\left[
\frac{1}{3}\left(2x^2+5x-1-\frac{6x^2\ln x}{x-1}\right)-\frac{2}{3}\left(x^2-5x-2+\frac{6x\ln x}{x-1}
\right)\right]\,,
\eea
with $M_u$ and $M_d$ standing for the masses of generic up- or down-type quarks respectively.

This form of $T$ is easy to understand. 
A helicity flip between incoming $\tau$ and outgoing lepton is needed, so both
the tensor and the pseudotensor amplitudes in eq.\ (\ref{Mrad})
will be proportional
to $M_\tau$ (assuming that $M_e$ and $M_\mu$ can be neglected). They will also
be proportional to the product of two RPV couplings, and the electric charge
$Q$ of the particle coupling to the photon. There will be a further factor of
$1/16\pi^2$ coming from the loop, and $1/{\tilde{M}}^2$ from the sparticle
propagator. We refer the reader to \cite{tau1} for the relevant Feynman 
diagrams. 

The decay width can be written as
\be
\Gamma(\tau\to \ell_i+\gamma) =
\frac{M_\tau^3}{4\pi} |T|^2 \,,
\ee
and the branching ratio is
\be
{\rm BR}(\tau\to\ell_i \gamma)=\frac{48\pi^2}{G_F^2 M_\tau^2}|T|^2\,.
\ee

%%%%%%%%%%%%%%

%-----------------------------------------
\begin{table}[htbp]
\begin{center}
\begin{tabular}{||c|c||c|c||}
\hline
Mode & BR (upper limit) & Mode & BR (upper limit) \\
     & ($\times 10^8$) &  & ($\times 10^8$)\\
\hline
$e \gamma$ & 3.3 & $\mu\gamma$ & 4.4\\
\hline
$e\pi^0$ & 2.2 & $\mu\pi^0$ & 2.7 \\
$e\eta $ & 4.4 & $\mu\eta$  & 2.3 \\
$e\eta'$ & 3.6 & $\mu\eta'$ & 3.8 \\
$e K_S $ & 2.6 & $\mu K_S$ & 2.3 \\
\hline
$e\rho^0$  & 1.8 & $\mu\rho^0$   & 1.2\\
$e\phi  $  & 3.1 & $\mu\phi  $   & 8.4\\
$e\omega$  & 4.8 & $\mu\omega$   & 4.7\\
$e K^{\ast 0}$  & 3.2 & $\mu K^{\ast 0}$    & 7.2\\
$e \overline{K^{\ast 0}}$  & 3.4 & $\mu \overline{K^{\ast 0}}$    & 7.0\\
\hline
$e^-e^+e^-$    & 2.7 & $\mu^-\mu^+\mu^-$    & 2.1 \\
$e^-e^+\mu^-$    & 1.8 & $\mu^-\mu^+  e^-$    & 2.7 \\
$e^-e^-\mu^+$    & 1.5 & $\mu^-\mu^-  e^+$    & 1.7 \\
\hline
\end{tabular}
\caption{Upper limits of various lepton-flavor violating $\tau$ decays.
The numbers are taken from: \cite{l-gamma} (radiative), \cite{l-p} (lepton
+ pseudoscalar), \cite{l-v} (lepton + vector), \cite{lll} (trilepton).}
\label{tab:data}
\end{center}
\end{table}

%-------------------------
The upper bounds at 90\% confidence limit (CL)
on the branching ratios of several
lepton-flavor violating (LFV) $\tau$ decay channels are shown in Table 
\ref{tab:data}. The quoted numbers are for integrated luminosities of
470 fb$^{-1}$ at $\Upsilon(4S)$ plus 46 fb$^{-1}$ at $\Upsilon(3S)$ and
$\Upsilon(2S)$
(SLAC, radiative decays), 802 fb$^{-1}$ at $\Upsilon(4S)$ and 99 fb$^{-1}$ at
$\Upsilon(5S)$ (KEK-B, lepton plus a pseudoscalar meson),  
782 fb$^{-1}$ at $\Upsilon(4S)$ and 72 fb$^{-1}$ at
$\Upsilon(5S)$ (KEK-B, lepton plus a vector meson), and 
782 fb$^{-1}$ at $\Upsilon(4S)$ (KEK-B, trilepton). It is useful to note that
apart from very few numbers coming from radiative decays with top quark
in the loop, the bounds on the product couplings are given by
\be
{\rm Actual~bound} = {\rm Bound~quoted}\times \left( \frac{\tilde{M}}{100}\right)
^2 \times \left( \frac{ {\rm Actual~BR~bound}}{{\rm Quoted~BR~bound}}\right)^{1/2}\,,
\label{bounds}
\ee
where the quoted bounds on branching ratios are those to be found in Table 1
while the actual bounds are the possibly improved numbers as can be obtained from 
LHC-B or Super B factories.  

%%%%%%%%%%%%%%%%%

\section{Results}

%%%%%%%%%%%%%%%%%%%%%%%
\begin{table}[htbp]
\begin{center}
\begin{tabular}{|c|c|c|c|}
\hline
$\lambda\lambda$ & Process & Our bound & Previous bound\\
\hline
(121)(123) $^{e\gamma}$ & $\tau\r ee\bar{e}$ & $1.8\times 10^{-4}$ & $7.0\times 10^{-4}$\\
(121)(123) $^\dag$ & $\tau\r e\mu\bar{\mu}$ & $2.6\times 10^{-4}$ & $7.0\times 10^{-4}$\\
(121)(131) $^{\nu 23, \mu\gamma}$ & $\tau\r\mu e\bar{e}$ & $2.1\times 10^{-4}$ & $6.8\times 10^{-4}$\\
(121)(132) $^{\nu 13}$ & $\tau\r \mu\mu\bar{e}$ & $1.4\times 10^{-4}$ & $5.6\times 10^{-4}$\\
(121)(231) $^{e\gamma}$ & $\tau\r ee\bar{e}$ & $1.8\times 10^{-4}$ & $7.0\times 10^{-4}$\\
(121)(232)& $\tau\r\mu e\bar{e}$ & $2.1\times 10^{-4}$ & $6.8\times 10^{-4}$\\
(122)(123) $^{\mu\gamma}$ & $\tau\r \mu\mu\bar{\mu}$ & $1.6\times 10^{-4}$ & $6.8\times 10^{-4}$\\
(122)(123) $^\dag$& $\tau\r\mu e\bar{e}$ & $2.1\times 10^{-4}$ & $6.8\times 10^{-4}$\\
(122)(131)& $\tau\r e\mu\bar{\mu}$ & $2.6\times 10^{-4}$ & $7.0\times 10^{-4}$\\
(122)(132) $^{\mu\gamma}$ & $\tau\r \mu\mu\bar{\mu}$ & $1.6\times 10^{-4}$ & $6.8\times 10^{-4}$\\
(122)(231)& $\tau\r ee\bar{\mu}$ & $1.4\times 10^{-4}$ & $5.2\times 10^{-4}$\\
(122)(232) $^{\nu 13, e\gamma}$ & $\tau\r e\mu\bar{\mu}$ & $2.6\times 10^{-4}$ & $7.0\times 10^{-4}$\\
(131)(133) $^{\nu 13, e\gamma}$ & $\tau\r ee\bar{e}$ & $1.8\times 10^{-4}$ & $7.0\times 10^{-4}$\\
(131)(233)& $\tau\r\mu e\bar{e}$ & $2.1\times 10^{-4}$ & $6.8\times 10^{-4}$\\
(132)(133) $^{\mu\gamma}$ & $\tau\r\mu e\bar{e}$ & $2.1\times 10^{-4}$ & $6.8\times 10^{-4}$\\
(132)(233) $^{\nu 13}$ & $\tau\r \mu\mu\bar{e}$ & $1.4\times 10^{-4}$ & $5.6\times 10^{-4}$\\
(133)(231) $^{\nu 23}$& $\tau\r ee\bar{\mu}$ & $1.4\times 10^{-4}$ & $5.2\times 10^{-4}$\\
(133)(232)& $\tau\r e\mu\bar{\mu}$ & $2.6\times 10^{-4}$ & $7.0\times 10^{-4}$\\
(231)(233) $^{e\gamma}$ & $\tau\r e\mu\bar{\mu}$ & $2.6\times 10^{-4}$ & $7.0\times 10^{-4}$\\
(232)(233) $^{\nu 23, \mu\gamma}$ & $\tau\r \mu\mu\bar{\mu}$ & $1.6\times 10^{-4}$ & $6.8\times 10^{-4}$\\
\hline
\end{tabular}

\caption {Bounds on $\lambda_{ijk}\lambda_{pqr}$  from  
$\tau\r \ell_i\ell_j\ell_k$ decays. All sleptons are assumed to be degenerate
at 100 GeV, see eq.\ (\ref{bounds}).
The superscript $\nu ij$ indicates that the
combination generates the $ij$-th entry of the neutrino mass matrix ${\cal M}_\nu$;
for more details, see text. The couplings marked with $e\gamma$ have a less severe 
upper bound of $1.2\times 10^{-2}$ coming from $\tau\to e+\gamma$; similarly,
those marked with $\mu\gamma$ have an upper bound of $1.4\times 10^{-2}$
from $\tau\to\mu+\gamma$. The bounds marked with a dagger are 
not the best bounds right now, but we keep them as they are of comparable order.}
\label{tab:trilepton}
\end{center}
\end{table}
%%%%%%%%%%%%%%%%%%%%

%%%%%%%%%%%%%%%%%%%%%%

\begin{table}[htbp]
\begin{center}
\begin{tabular}{|c|c|c|c|}
\hline
$\lambda\lambda'$ & Process & Our bound & Previous bound\\
\hline
(123)(111)& $\tau\r \mu\eta$ & $3.2\times 10^{-5}$ & $6.7\times 10^{-5}$\\
(123)(112)& $\tau\r \mu K_S$ & $1.4\times 10^{-4}$ & $1.0\times 10^{-3}$\\
(123)(121)& $\tau\r \mu K_S$ & $1.4\times 10^{-4}$ & $1.0\times 10^{-3}$\\
(123)(122)& $\tau\r \mu\eta$ & $3.1\times 10^{-4}$ & $3.7\times 10^{-4}$\\
(123)(211)& $\tau\r e\eta$ & $4.5\times 10^{-5}$ & $8.5\times 10^{-5}$\\
(123)(212)& $\tau\r e K_S$ & $1.5\times 10^{-4}$ & $9.7\times 10^{-4}$\\
(123)(221)& $\tau\r e K_S$ & $1.5\times 10^{-4}$ & $9.7\times 10^{-4}$\\
(123)(222)& $\tau\r e\eta$ & $4.3\times 10^{-4}$ & $4.6\times 10^{-4}$\\
(131)(111)& $\tau\r e\eta$ & $4.5\times 10^{-5}$ & $8.5\times 10^{-5}$\\
(131)(112)& $\tau\r e K_S$ & $1.5\times 10^{-4}$ & $9.7\times 10^{-4}$\\
(131)(121)& $\tau\r e K_S$ & $1.5\times 10^{-4}$ & $9.7\times 10^{-4}$\\
(131)(122)& $\tau\r e\eta$ & $4.3\times 10^{-4}$ & $4.6\times 10^{-4}$\\
(132)(111)& $\tau\r \mu\eta$ & $3.2\times 10^{-5}$ & $6.7\times 10^{-5}$\\
(132)(112)& $\tau\r \mu K_S$ & $1.4\times 10^{-4}$ & $1.0\times 10^{-3}$\\
(132)(121)& $\tau\r \mu K_S$ & $1.4\times 10^{-4}$ & $1.0\times 10^{-3}$\\
(132)(122)& $\tau\r \mu\eta$ & $3.1\times 10^{-4}$ & $3.7\times 10^{-4}$\\
(133)(311)& $\tau\r e\eta$ & $4.5\times 10^{-5}$ & $8.5\times 10^{-5}$\\
(133)(312)& $\tau\r e K_S$ & $1.5\times 10^{-4}$ & $9.7\times 10^{-4}$\\
(133)(321)& $\tau\r e K_S$ & $1.5\times 10^{-4}$ & $9.7\times 10^{-4}$\\
(133)(322)& $\tau\r e\eta$ & $4.3\times 10^{-4}$ & $4.6\times 10^{-4}$\\
(231)(211)& $\tau\r e\eta$ & $4.5\times 10^{-5}$ & $8.5\times 10^{-5}$\\
(231)(212)& $\tau\r e K_S$ & $1.5\times 10^{-4}$ & $9.7\times 10^{-4}$\\
(231)(221)& $\tau\r e K_S$ & $1.5\times 10^{-4}$ & $9.7\times 10^{-4}$\\
(231)(222)& $\tau\r e\eta$ & $4.3\times 10^{-4}$ & $4.6\times 10^{-4}$\\
(232)(211)& $\tau\r \mu\eta$ & $3.2\times 10^{-5}$ & $6.7\times 10^{-5}$\\
(232)(212)& $\tau\r \mu K_S$ & $1.4\times 10^{-4}$ & $1.0\times 10^{-3}$\\
(232)(221)& $\tau\r \mu K_S$ & $1.4\times 10^{-4}$ & $1.0\times 10^{-3}$\\
(232)(222)& $\tau\r \mu\eta$ & $3.1\times 10^{-4}$ & $3.7\times 10^{-4}$\\
(233)(311)& $\tau\r \mu\eta$ & $3.2\times 10^{-5}$ & $6.7\times 10^{-5}$\\
(233)(312)& $\tau\r \mu K_S$ & $1.4\times 10^{-4}$ & $1.0\times 10^{-3}$\\
(233)(321)& $\tau\r \mu K_S$ & $1.4\times 10^{-4}$ & $1.0\times 10^{-3}$\\
(233)(322)& $\tau\r \mu\eta$ & $3.1\times 10^{-4}$ & $3.7\times 10^{-4}$\\
\hline
\end{tabular}
\caption {Bounds on $\lambda_{ijk}\lambda'_{pqr}$  from  
$\tau\r \ell + M$ decays, where $M$ is a pseudoscalar meson. Again, all
sleptons are assumed degenerate at 100 GeV.}
\label{tab:meson1}
\end{center}
\end{table}

\begin{table}[htbp]
\begin{center}
\begin{tabular}{|c|c|c|c|}
\hline
$\lambda'\lambda'$ & Process & Our bound & Previous bound\\
\hline
(111)(311) $^{\nu 11}$ & $\tau\r e\rho$ & $2.3\times 10^{-4}$ & $2.4\times 10^{-3}$\\
(111)(312)& $\tau\r e\ksb$ & $2.9\times 10^{-4}$ & $3.6\times 10^{-3}$\\
(112)(311)& $\tau\r e\kst$ & $2.8\times 10^{-4}$ & $2.9\times 10^{-3}$\\
(112)(312)& $\tau\r e\phi$ & $2.3\times 10^{-4}$ & $3.4\times 10^{-3}$\\
(113)(313)& $\tau\r e\rho$ & $2.3\times 10^{-4}$ & $2.4\times 10^{-3}$\\
(121)(321)& $\tau\r e\rho$ & $2.3\times 10^{-4}$ & $2.4\times 10^{-3}$\\
(121)(322)& $\tau\r e\ksb$ & $2.9\times 10^{-4}$ & $3.6\times 10^{-3}$\\
(122)(321)& $\tau\r e\kst$ & $2.8\times 10^{-4}$ & $2.9\times 10^{-3}$\\
(122)(322) $^{\nu 22}$ & $\tau\r e\phi$ & $2.3\times 10^{-4}$ & $3.4\times 10^{-3}$\\
(131)(331)& $\tau\r e\rho$ & $2.3\times 10^{-4}$ & $2.4\times 10^{-3}$\\
(131)(332)& $\tau\r e\ksb$ & $2.9\times 10^{-4}$ & $3.6\times 10^{-3}$\\
(132)(331)& $\tau\r e\kst$ & $2.8\times 10^{-4}$ & $2.9\times 10^{-3}$\\
(132)(332)& $\tau\r e\phi$ & $2.3\times 10^{-4}$ & $3.4\times 10^{-3}$\\
(211)(311) $^{\nu 11}$ & $\tau\r \mu\rho$ & $1.9\times 10^{-4}$ & $4.3\times 10^{-3}$\\
(211)(312)& $\tau\r \mu K_S$ & $3.8\times 10^{-4}$ & $2.4\times 10^{-3}$\\
(212)(311)& $\tau\r \mu\kst$ & $3.8\times 10^{-4}$ & $3.6\times 10^{-3}$\\
(212)(312)& $\tau\r \mu\rho$ & $1.9\times 10^{-4}$ & $4.3\times 10^{-3}$\\
(213)(313)& $\tau\r \mu\rho$ & $1.9\times 10^{-4}$ & $4.3\times 10^{-3}$\\
(221)(321)& $\tau\r \mu\rho$ & $1.9\times 10^{-4}$ & $4.3\times 10^{-3}$\\
(222)(321)& $\tau\r \mu\kst$ & $3.8\times 10^{-4}$ & $3.6\times 10^{-3}$\\
(222)(322) $^{\nu 22}$ & $\tau\r \mu\phi$ & $3.8\times 10^{-4}$ & $3.4\times 10^{-3}$\\
(231)(331)& $\tau\r \mu\rho$ & $1.9\times 10^{-4}$ & $4.3\times 10^{-3}$\\
(231)(332)& $\tau\r \mu K_S$ & $3.8\times 10^{-4}$ & $2.4\times 10^{-3}$\\
(232)(331)& $\tau\r \mu\kst$ & $3.8\times 10^{-4}$ & $3.6\times 10^{-3}$\\
(232)(332)& $\tau\r \mu\phi$ & $3.8\times 10^{-4}$ & $3.4\times 10^{-3}$\\
\hline
\end{tabular}
\caption 
{Bounds on $\lambda'_{ijk}\lambda'_{pqr}$  from  
$\tau\r \ell + M$ decays, where $M$ is a pseudoscalar or
a vector meson. All
squarks are assumed degenerate at 100 GeV. The superscript ${\nu ii}$
indicates that the combination contributes to the $ii$-th element of the
neutrino mass matrix.}
\label{tab:meson2}
\end{center}
\end{table}

%%%%%%%%%%%%%%%%%%%%

%%%%%%%%%%%%%%%%%%%%

\begin{table}[htbp]
\begin{center}
 \begin{tabular}{|c|c||c|c|}
\hline
$\lambda'\lambda'$ & Upper bound & $\lambda'\lambda'$ & Upper bound\\
\hline
(111)(311)& $4.1\times 10^{-3}$ & (211)(311)& $4.7\times 10^{-3}$ \\
(112)(312)& $4.1\times 10^{-3}$ & (212)(312)& $4.7\times 10^{-3}$\\
(113)(313)& $4.1\times 10^{-3}$ & (213)(313)& $4.8\times 10^{-3}$\\
(121)(321)& $4.1\times 10^{-3}$ & (221)(321)& $4.7\times 10^{-3}$\\
(122)(322)& $4.1\times 10^{-3}$ & (222)(322)& $4.7\times 10^{-3}$\\
(123)(323)& $4.1\times 10^{-3}$ & (223)(323)& $4.8\times 10^{-3}$\\
(131)(331)& $7.8\times 10^{-2}$ & (231)(331)& $8.9\times 10^{-2}$\\
(132)(332)& $7.8\times 10^{-2}$ & (232)(332)& $8.9\times 10^{-2}$\\
(133)(333)& $10.4\times 10^{-2}$& (233)(333)& $12.0\times 10^{-2}$\\
\hline
\end{tabular}
\caption {Bounds on $\lambda'\lambda'$ type products from $\tau\r e +\gamma$ (left)
and $\tau\r\mu+\gamma$ (right) decays.}
\label{tab:rad2}
\end{center}
\end{table}

%----------

Our results are shown in Tables \ref{tab:trilepton}-\ref{tab:rad2}.
Table \ref{tab:trilepton} shows the bounds on $|\lambda_{ijk}\lambda_{ilm}|$
as obtained from the lepton flavour violating trileptonic $\tau$ decays, see eqs.\
(\ref{4fer}) and (\ref{labc}), taking into account the fact that $\lambda_{ijk}$
is antisymmetric in $i$ and $j$. The last column in each of the Tables 
\ref{tab:trilepton}-\ref{tab:meson2} is taken from 
\cite{tau4}. 

Tables \ref{tab:meson1} and \ref{tab:meson2} show the bounds on $\lambda
\lambda'$ and $\lambda'\lambda'$ type products (again, magnitudes only,
as all these processes are SM-forbidden and hence single-amplitude processes)
coming from $\tau$ decays to a charged lepton and a neutral meson. 

Neutrino masses put the tightest constraint on some of the RPV couplings. A typical
bound from the diagonal entries of the neutrino mass matrix ${\cal M}_\nu$ is
\cite{dreiner2}
\be
|\lambda'_{i11}| < 6.0\times 10^{-3}\,,\ \ 
|\lambda'_{i22}| < 3.5\times 10^{-5}\,,\ \ 
|\lambda'_{i33}| < 8.9\times 10^{-6}\,.
\ee

Assuming all sleptons to be degenerate (and all squarks too), and the trilinear $A$-terms
to be proportional to the corresponding Yukawa couplings, the lepton-slepton and
quark-squark loops contribute to the neutrino mass matrix, even in the absence of any
bilinear RPV interaction:
\bea
\left[ {\cal M}_{\nu} \right]_{ij} &\approx& \frac{1}{8\pi^2 {\tilde{M}_e}^2} (A^e-\mu\tan\beta)
\sum_{k,l} \lambda_{ikl}\lambda_{jlk} M_{e_k} M_{e_l}\,,\nonumber\\
 \left[ {\cal M}_{\nu} \right]_{ij} &\approx& \frac{3}{8\pi^2 {\tilde{M}_d}^2} (A^d-\mu\tan\beta)
\sum_{k,l} \lambda'_{ikl}\lambda'_{jlk} M_{d_k} M_{d_l}\,.
\label{numass}
\eea
The combinations that affect eq.\ (\ref{numass}) are shown with superscripts $\nu ij$ in the
respective tables.   

A total of 12 $\lambda\lambda$ combinations are bound from the radiative decays $\tau\to
e+\gamma$ and $\tau\to\mu+\gamma$. Among them, ten have been shown in Table 2;
they are definitely less severe than those coming from trilepton decays. We also have
\be
|\lambda_{123}\lambda_{233}| < 1.2\times 10^{-2} ~~ (e+\gamma)\,,\ \ 
|\lambda_{123}\lambda_{133}| < 1.4\times 10^{-2} ~~ (\mu+\gamma)\,.
\ee

Table \ref{tab:rad2} shows the bounds on $\lambda'\lambda'$ type
products
coming from radiative decays $\tau\r\mu\gamma$ and $\tau\r e\gamma$. 
Apart from the combinations $\lambda'_{i23}\lambda'_{323}$ and $\lambda'_{i33}
\lambda'_{333}$ ($i,j=1,2$), all other combinations have a better bound as shown in 
Table \ref{tab:meson2}.

\section{Summary and conclusion}

We have found limits on several product couplings of types $\lambda\lambda$,
$\lambda\lambda'$, and $\lambda'\lambda'$ coming from lepton flavour violating
$\tau$ decays. These bounds should be marginally improved once the Belle collaboration
finishes its data analysis. However, any super B factory will do a lot better, and 
these bounds can go up by orders of magnitude. 

While some of these products do affect the neutrino masses and hence are possibly
more tightly constrained than that discussed here, there are other low-energy processes
that might be affected by these couplings. For example, the decay $K\to\pi\nu\bar\nu$,
which is supposed to be a clean channel for the determination of the CP violating phase
$\sin(2\beta)$, gets contribution from couplings like $\lambda'_{i2k}\lambda'_{j1k}$
or $\lambda'_{ik1}\lambda'_{jk2}$.  

Another interesting prospect is to find these LFV decays at the LHC. For the $\lambda\lambda'_{i11}$
type couplings, one can observe a Drell-Yan type unlike-flavour dilepton production, mediated by
a slepton propagator. The same applies for $\lambda'\lambda'$ type couplings where the first
generation quark fields come into play. These signals will be interesting to study.

\centerline{\bf{Acknowledgement}}
The author acknowledges the University Grants Commission, Government of India, for financial support. She also acknowledges helpful discussions with A. Kundu.

\end{document}